\documentclass[twocolumn,amsmath,amssymb,footinbib,superscriptaddress]{revtex4-1}

\usepackage{graphicx}
\usepackage{dcolumn}
\usepackage{bm}
\usepackage{color}
\usepackage{amsmath}
\usepackage[colorlinks=true,citecolor=blue,urlcolor=blue,linkcolor=blue]{hyperref}

\begin{document}

\title{Coherent optical and acoustic phonons generated at lattice-matched \\GaP/Si(001) heterointerfaces}

\author{Kunie Ishioka}
\email{ishioka.kunie@nims.go.jp}
\affiliation{National Institute for Materials Science, Tsukuba, 305-0047 Japan}

\author{Andreas Beyer}
\affiliation{Faculty of Physics and Materials Sciences Center, Philipps-Universit{\"a}t Marburg, 35032 Marburg, Germany}

\author{Wolfgang Stolz}
\affiliation{Faculty of Physics and Materials Sciences Center, Philipps-Universit{\"a}t Marburg, 35032 Marburg, Germany}

\author{Kerstin Volz}
\affiliation{Faculty of Physics and Materials Sciences Center, Philipps-Universit{\"a}t Marburg, 35032 Marburg, Germany}

\author{Hrvoje Petek}
\affiliation{Department of Physics and Astronomy, University of Pittsburgh, Pittsburgh, PA 15260, USA}

\author{Ulrich H{\"o}fer}
\affiliation{Faculty of Physics and Materials Sciences Center, Philipps-Universit{\"a}t Marburg, 35032 Marburg, Germany}

\author{Christopher J. Stanton}
\affiliation{Department of Physics, University of Florida, Gainesville, FL 32611 USA}

\date{\today}

\begin{abstract}
Thin GaP films can be grown on an exact Si(001) substrate with nearly perfect lattice match.  We perform all-optical pump-probe measurements to investigate the ultrafast electron-phonon coupling at the buried interface of GaP/Si.  Above-bandgap excitation with a femtosecond laser pulse can induce coherent longitudinal optical (LO) phonons both in the GaP opverlayer and in the Si substrate.  The coupling of the GaP LO phonons with photoexcited plasma is reduced significantly with decreasing the GaP layer thickness from 56 to 16 nm due to the quasi-two-dimensional confinement of the plasma.  The same laser pulse can also generate coherent longitudinal acoustic (LA) phonons in the form of a strain pulse.  The strain induces not only a periodic modulation in the optical reflectivity as they propagate in the semiconductors, but also a sharp spike when it arrives at the GaP layer boundaries.  The acoustic pulse induced at the GaP/Si interface is remarkably stronger than that at the Si surface, suggesting a possible application of the GaP/Si heterostructure as an opto-acoustic transducer.  The amplitude and the phase of the reflectivity modulation varies with the GaP layer thickness, which can be understood in terms of the interference caused by the  multiple acoustic pulses generated at the top surface and at the buried interface.
\end{abstract}

\maketitle

\section{INTRODUCTION}

Internal interfaces between two semiconductors are central to most modern micro-electronic devices, and the characterization of the electronic states at such interfaces has been one of the most pressing issues in device physics \cite{MonchBook}. Unfortunately, angle-resolved photoelectron spectroscopy (ARPES), the standard technique to directly map out the band structure at surfaces and in thin solid films, generally lacks the probing depth to access buried interfaces.  All-optical spectroscopies that do not suffer from this problem thus play an important role for an improved microscopic understanding of buried semiconductor heterointerfaces.  A promising model system to develop and test the new spectroscopic approaches is GaP/Si(001), with gallium phosphide (GaP) and silicon (Si) having a nearly perfect lattice match. Whereas both materials are indirect-gap semiconductors with the conduction band minimum at the $X$ point, their bandgap energies are considerably different, making the band alignment between the two semiconductor a target of extensive experimental and theoretical studies \cite{Zeidenbergs1967, Katoda1980, Dandrea1990, Sakata2008, Pal2013}.  
Recently GaP layers free from dislocations, staking faults, or twins were successfully grown directly on exact Si(001) substrate.  Recently GaP layers free from dislocations, staking faults, or twins were successfully grown directly on exact Si(001) substrate \cite{Lin2013, Supplie2014, Volz2011, Beyer2011, Beyer2012}.  Atomically-resolved transmission electron microscopy studies revealed that the interface consists of a  pyramidal structure with intermixing of $\sim$7 atomic layers \cite{Beyer2016}. The well-defined GaP/Si heterointerface has the potential for application in Si-based optoelectronic devices and high efficiency multi-junction solar cells \cite{Wagner2014, Saive2018}, as well as in integrating dilute nitride mixed compound Ga(NAsP) that has a direct band structure and lasing operation ability on the Si substrate \cite{Kunert2006, Liebich2011}.

In our previous study \cite{Ishioka2016} we reported on an all-optical approach to evaluate electronic band structure at the buried GaP/Si(001) interfaces by means of coherent phonon spectroscopy.  The technique is based on the generation of coherent longitudinal optical (LO) phonons of a polar semiconductor via transient screening of the depletion field (TDFS) mechanism, in which ultrafast drift-diffusion current suddenly screens the built-in electric field in the depletion region at the surface or interface  \cite{DekorsyBook, Foerst2007, Ishioka2015}.  We performed pump-probe reflectivity measurements using 10-fs near ultraviolet (NUV) pulses on thin GaP films grown on Si(001) under different conditions, and successfully reconstructed the band bendings of the heterostructures from the experimentally observed coherent phonon amplitudes.    The same pump-probe scheme was also used to investigate coherent longitudinal acoustic (LA) phonons in the GaP/Si heterostructures  \cite{Ishioka2017APL}.    
Because of the short penetration depth of the NUV \emph{probe} pulse, however, we could monitor the acoustic phonon dynamics for only a few tens of picoseconds.  To obtain a more complete and quantitative understanding we need to follow the propagation of the acoustic phonons in Si over much longer temporal and deeper spatial scales.
	
In semiconductor heterostructures,  coherent optical and acoustic phonons often exhibit different features from those in bulk semiconductors.  In GaAs/AlGaAs quantum wells, for example, the Coulomb interaction between the LO phonons and the electron plasma, which would give rise to the LO phonon-plasmon coupled (LOPC) mode for the bulk GaAs, becomes weaker with decreasing GaAs well width due to the reduced spectral overlap between electronic and phononic bands 
\cite{Dekorsy1996}.  When the Bloch oscillation energy in GaAs/AlGaAs superlattices  \cite{Dekorsy2000} or the splitting energy of the heavy- and light-hole excitons in GaAs/AlAs multiple quantum wells (MQWs)  \cite{Mizoguchi2004} is tuned into resonance with the GaAs LO phonon energy, by contrast, the coherent electronic wavefunctions can drive the coherent phonons strongly.    
In GaAs/AlAs superlattices, the backfolding of the bulk phonon dispersion into the mini Brillouin zones leads to the emergence of coherent zone-folded acoustic phonons \cite{Yamamoto1994, Bartels1999, Mizoguchi2002}.  Photoexcitation of an InGaN/GaN MQW by a single femtosecond NUV pulse can generate more intense coherent acoustic phonons 
\cite{Sun1999, Sun2000, Liu2005, Wen2009}  than those excited in a GaN film using a transient grating technique \cite{Huang2001}.  
In contrast to these extensive studies on the GaAs- and GaN-based heterostructures, the electron-phonon couplings in heterostructures based on other semiconductors have not been investigated systematically  \cite{Takeuchi2001, Lim2003, Noe2012, Wang2005}.	
	
In the present study we investigate the electron-phonon coupling dynamics at the GaP/Si heterointerfaces with different GaP overlayer thicknesses up to $\sim$60 nm.  We perform one-color pump-probe reflectivity measurements with NUV pump and probe pulses to monitor the coherent optical and acoustic phonons with 10-fs time resolution over a few tens of ps, and two-color measurements using NUV pump and visible probe to follow the propagation of the acoustic phonons in Si on sub-nanosecond time scale.  We find that the GaP LO phonon-plasmon coupling depends critically on the GaP layer thickness due to the two-dimensional confinement of the plasmons in the thinner GaP layers.  The amplitude of the sub-THz modulation in the transient reflectivity induced by the propagating coherent acoustic phonons also strongly depends on the GaP layer thickness, which can be explained in terms of the interference between reflections by multiple acoustic pulses generated at the GaP top surface and at the GaP/Si heterointerface and propagating into Si. 

\section{EXPERIMENTAL METHODS}

The samples studied are GaP thin films grown epitaxially on Si (001) substrates by metal organic vapor phase epitaxy (MOVPE).  Details of the MOVPE growth and the structural characterization were described elsewhere \cite{Volz2011,Beyer2011,Beyer2012}.   (001)-oriented Si wafers with a 0.1$^\circ$ intentional miscut in the [110] direction are used to support the formation of double-steps.  After a wet chemical cleaning procedure and a high-temperature heat treatment of the substrate, a Si buffer layer is grown in a chemical vapor deposition (CVD) process using silane.  The GaP-layer is then grown on the buffer layer.  The first nucleation step is a pulsed growth scheme, where the precursors for Ga and P are offered intermittently at 450$^\circ$C.  This is followed by continuous GaP overgrowth, in which both precursors are injected simultaneously at 675$^\circ$C, for different temporal durations to achieve different overlayer thicknesses from 16 to 56 nm.  Structural characterization and polarity determination of the GaP/Si samples are performed by transmission electron microscope (TEM).  Cross sectional images of the interfaces confirm the absence of major dislocations, stacking faults or twins in the GaP overlayers and the formation of self-annihilating anti-phase domains (APDs).  For comparison, (001)-oriented $n$-doped Si and GaP single crystal wafers are also investigated.

\begin{figure}
\includegraphics[
width=0.45\textwidth]{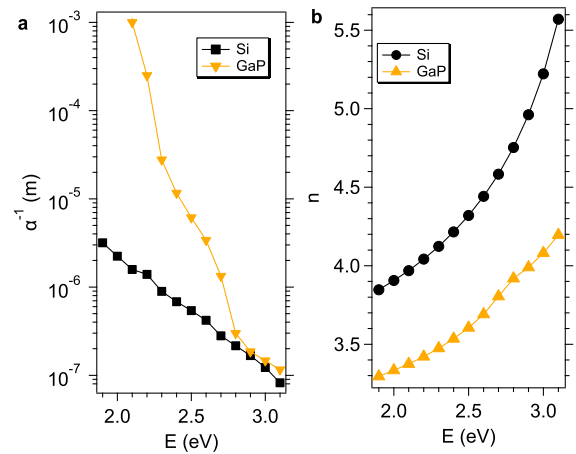}
\caption{\label{alpha_n} Penetration depth $\alpha^{-1}$ (\textbf{a}) and refractive index $n$ (\textbf{b}) of Si and GaP as a function of photon energy $E$ \cite{Aspnes1983}. 
}
\end{figure}

Pump-probe reflectivity measurements are performed in a near back-reflection configuration under ambient conditions.  For the one-color measurements, the second harmonic of a Ti:sapphire oscillator with 3.1-eV photon energy (400-nm wavelength), 10-fs duration and 80-MHz repetition rate is used for both pump and probe pulses.  The NUV  photons can excite carriers across the direct band gap at the $\Gamma$ point of GaP and along the $L$ valleys of Si  \cite{Adachi1987, Adachi1988, Hase2003, Ishioka2017PRB}.    The optical penetration depths for the 3.1 eV light are $\alpha^{-1}$=116 nm and 82 nm in GaP and Si \cite{Aspnes1983}.  The pump and probe laser spots on the sample are $\sim$30  $\mu$m in diameter.  The pump-induced change in the reflectivity $\Delta R$ is measured as a function of time delay $t$ between pump and probe pulses using a fast scan technique.  This scheme allows us to monitor the the optical and acoustic phonons as well as the electronic response for the first few tens of ps with 10 fs time resolution.   

\begin{figure}
\includegraphics[
width=0.475\textwidth]{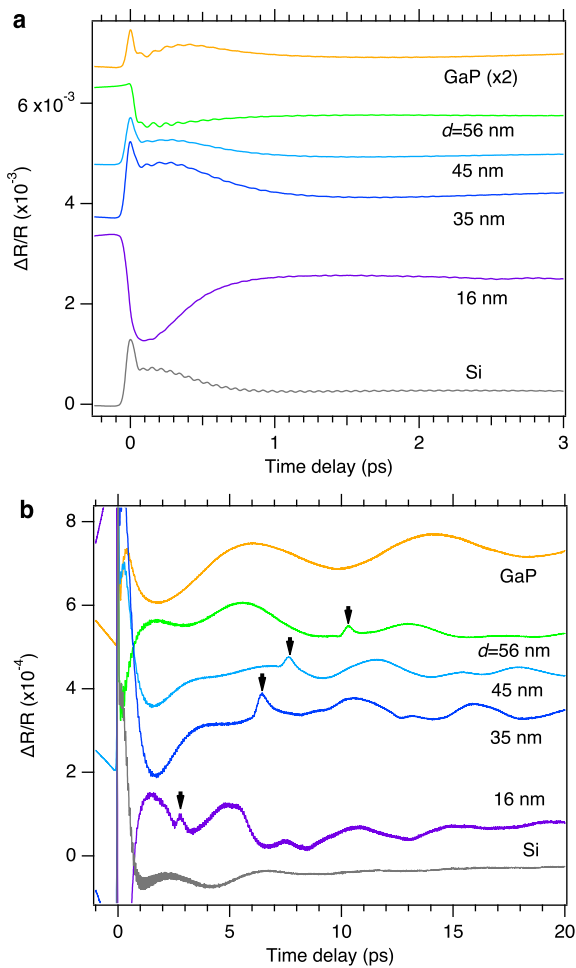}
\caption{\label{TD400nm} Transient reflectivity traces of GaP/Si(001) heterointerfaces with different GaP overlayer thicknesses $d$, together with those of bulk Si and GaP, pumped and probed at 3.1 eV.  Pump densities are 70 $\mu$J/cm$^2$ for bulk Si and 40 $\mu$J/cm$^2$ for other samples.  \textbf{a} and \textbf{b} show the same traces in different horizontal and vertical scales.  Arrows in \textbf{b} indicate the reflectivity spikes.  Traces are offset for clarity.
}
\end{figure}

For the two-color measurements, the second harmonic of a  Ti:sapphire regenerative amplifier output, with 3.1-eV photon energy, 150-fs duration and 100-kHz repetition rate, is used as the pump pulse, whereas the output of an optical parametric amplifier with tunable photon energy between 2.56 and 2.00 eV (wavelength between 485 and 620 nm) serves as the probe.  The pump and probe laser spots on the sample are $\sim$250 and $\sim180 \mu$m in diameter.  $\Delta R$ is measured as a function of $t$ using a slow scan technique.  This scheme allows us to monitor the acoustic pulses  on the spacial and temporal scales up to $\mu$m and ns, since the penetration depth of the visible probe light, summarized in Fig.~\ref{alpha_n}a,  is much larger than that of the NUV pump.

\section{RESULTS}
\subsection{One-color pump-probe measurements}

\begin{figure}
\includegraphics[
width=0.475\textwidth]{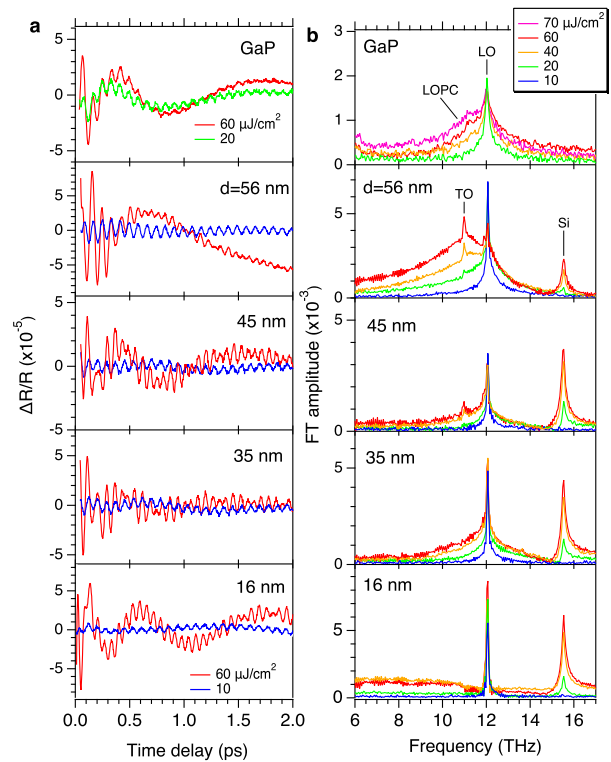}
\caption{\label{CFTC400nm} Oscillatory parts of the transient reflectivity changes (\textbf{a}) and their Fourier-transformed (FT) spectra (\textbf{b}) for GaP/Si(001) heterointerfaces with different GaP overlayer thicknesses $d$, together with those of bulk GaP,  pumped and probed at 3.1 eV.  Pump densities are varied from 10 to 60 $\mu$J/cm$^2$ for the GaP/Si samples and from 20 to 70 $\mu$J/cm$^2$ for bulk GaP.
}
\end{figure}

Figure~\ref{TD400nm} compares the pump-induced reflectivity changes $\Delta R/R$ of the GaP films on Si, together with those of the bulk Si and GaP, obtained in the one-color pump-probe scheme.  The traces for the bulk Si and GaP show a sharp spike at $t$=0, followed by an increase and then by a decay within 1 ps, as shown in Fig.~\ref{TD400nm}a.  The non-oscillatory responses provide information on the photoexcited carrier dynamics.  The  time constants of the increase and decay, 150 and 200 fs for Si, fall within the time scale of the carrier-phonon energy relaxation in the $L$ valley reported in the previous studies \cite{Sjodin1998, Sabbah2002, Ichibayashi2009, Ichibayashi2011,Sangalli2015, Meng2015}.  The time constants for GaP, 400 and 440 fs, correspond to the $\Gamma\rightarrow X_1$ and $\Gamma\rightarrow L$ scatterings \cite{Sjakste2007, Collier2013}.  The non-oscillatory reflectivity changes of the GaP/Si heterointerfaces cannot be described by simple linear combinations of the two bulk signals. 
For $d$=16 and 56 nm, especially, $\Delta R/R$ drops instead of rises after photoexcitation without exhibiting an initial spike.  The results suggest the importance of the interfacial carrier dynamics, such as the charge transfer across the heterointerface, in the reflectivity responses of the GaP/Si samples.  

The reflectivity traces are also modulated by coherent \emph{optical} phonons of GaP and Si with periods below 100 fs, as shown in Fig.~\ref{CFTC400nm}a.  Fourier-transformed (FT) spectra for \emph{bulk} GaP, shown in Fig.~\ref{CFTC400nm}b, exhibit a sharp peak of the LO mode at 12 THz at a low pump density of 10 $\mu$J/cm$^2$.  With increasing pump density, a broad band due to the LO phonon-plasmon coupled (LOPC) mode grows at a lower frequency, and eventually overwhelms the LO peak, indicating the almost complete screening of the LO mode at high pump densities of $\geq60 \mu$J/cm$^2$.  The frequency and the dephasing rate of the LOPC mode suggests that photoexcited mixed electron-hole plasma is responsible for the screening \cite{Ishioka2015}.

\begin{figure}
\includegraphics[
width=0.475\textwidth]{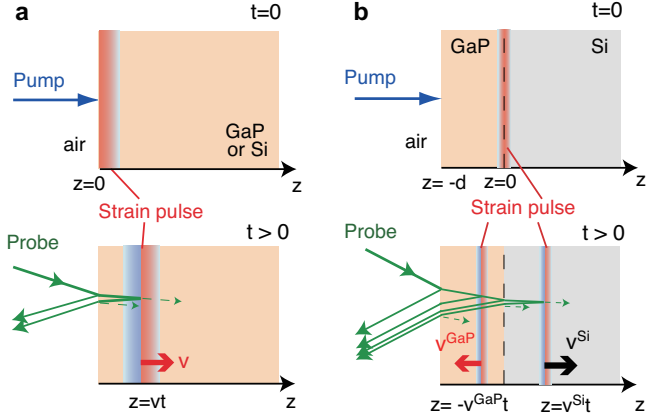}
\caption{\label{Scheme}  Schematics of the generation and optical detection of acoustic pulses at a semiconductor surface (\textbf{a}) and at a buried interface (\textbf{b}).  At $t=0$, pump light generates an acoustic pulse at the surface or interface (upper panels).  At $t>0$, probe light is reflected by the acoustic pulses propagating in the semiconductors in the depth direction as well as by the fixed surface and interface.  The interference between the reflected probe lights leads to the periodic modulation of the reflected probe intensity as a function of time delay between pump and probe.  The incident angle of the probe light is exaggerated for clarity.
}
\end{figure}

The GaP LO and LOPC modes are also observed in the GaP/Si heterointerface samples, together with the Si optical phonon at 15.6 THz.  The appearance of the LOPC mode in the FT spectra (Fig.~\ref{CFTC400nm}b) depends critically on the GaP layer thickness $d$, however.  At $d$=56 nm, the broad LOPC mode becomes dominant with increasing pump power, indicating as efficient screening of the LO phonons as in the bulk GaP.  At $d$=45 and 35 nm, by contrast, the growth of the LOPC mode with pump power is less efficient, leaving the LO mode partially unscreened even at the highest pump density.  At $d$=16 nm, the sharp LO mode dominates the FT spectra at all pump densities without exhibiting clear indication of the LOPC mode.    Our observation is consistent with the previous report on the well width-dependent LO-plasmon coupling in GaAs MQWs \cite{Dekorsy1996} and can essentially be explained in terms of the dimensionality-dependent dispersion relations of the plasmons.  For three-dimensional (3D)  and two-dimensional (2D) plasmons, the dispersion relations in the long wavelength limit are given by \cite{Schaich1992, Pines1956, Stern1967}: 
\begin{eqnarray}
\omega^{3D}(q)&=&\sqrt{\omega_{pl}^2+\dfrac{q^2}{c^2}};\\
\omega^{2D}(q)&=&\sqrt{\dfrac{\omega_{pl}^2q}{2}},
\end{eqnarray}
where $q$ is the wavevector and $\omega_{pl}=\sqrt{4\pi Ne/m^*\epsilon_\infty}$ is the plasma frequency as a function of carrier density $N$, effective mass $m^*$ and the high-frequency dielectric constant $\epsilon_\infty$. The 3D plasmons have nonzero frequency at $q\rightarrow0$, and can strongly couple with the LO phonons at $q\sim0$ when $\omega_{pl}$ is comparable to the LO phonon frequency \cite{Ishioka2015}.  By contrast, purely 2D plasmons have a  vanishing frequency at $q\rightarrow0$ and therefore would not interact with the LO phonons.    In the present study, the efficient (inefficient) screening of the LO phonons at $d$=56 nm (16 nm) indicates that the plasmons in the GaP film acquire a quasi-3D (quas-2D) dispersion relation; at the intermediate thicknesses the plasmon dispersion is inbetween.


\begin{figure}
\includegraphics[
width=0.475\textwidth]{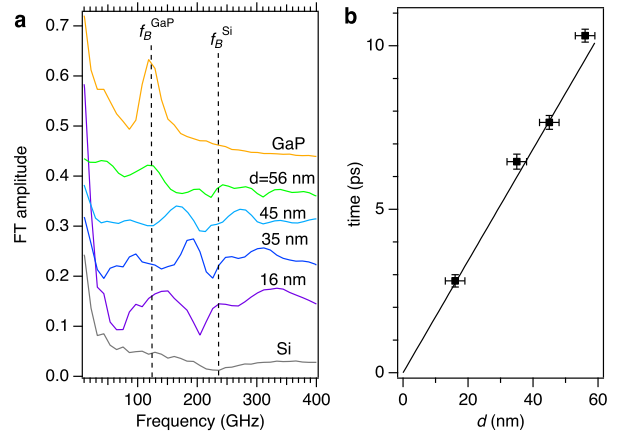}
\caption{\label{FT400nmL} \textbf{a}: FT spectra of the reflectivity changes shown in Fig.~\ref{TD400nm}b for time delay $t\geq$0.3 ps.  \textbf{b}: Appearance time of the reflectivity spike, marked by arrows in Fig.~\ref{TD400nm}b, as a function of the GaP film thickness $d$.  Solid line represents $t=d/v^\textrm{GaP}$. 
}
\end{figure}

\begin{figure*}
\includegraphics[
clip,width=0.9\textwidth]{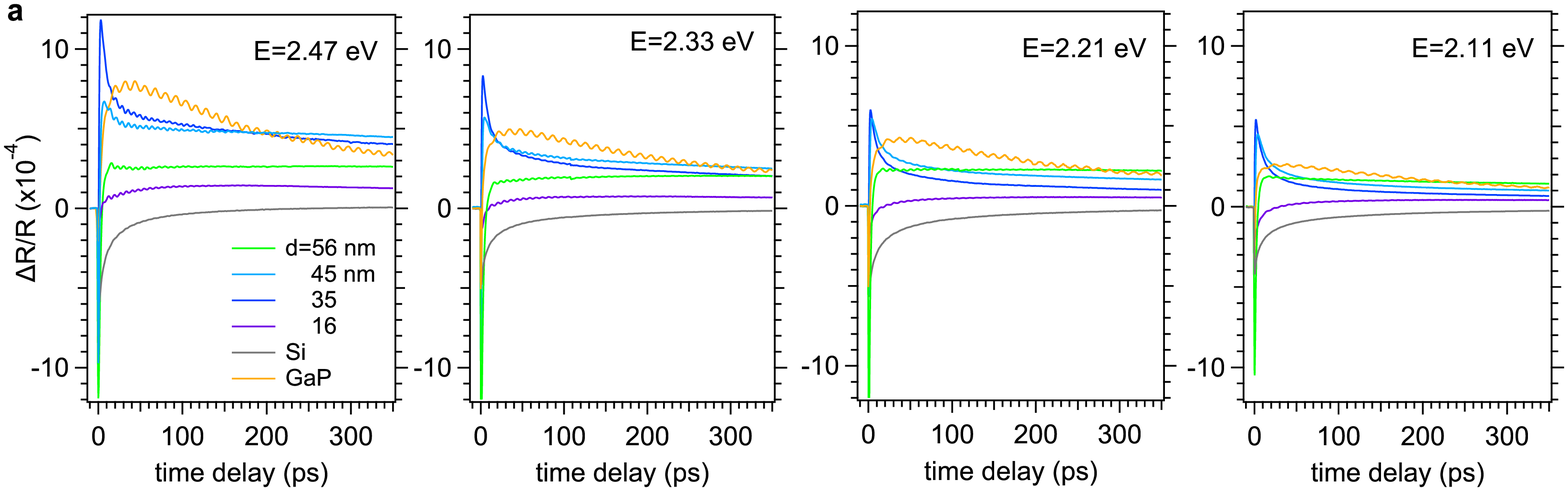}
\includegraphics[
width=0.9\textwidth]{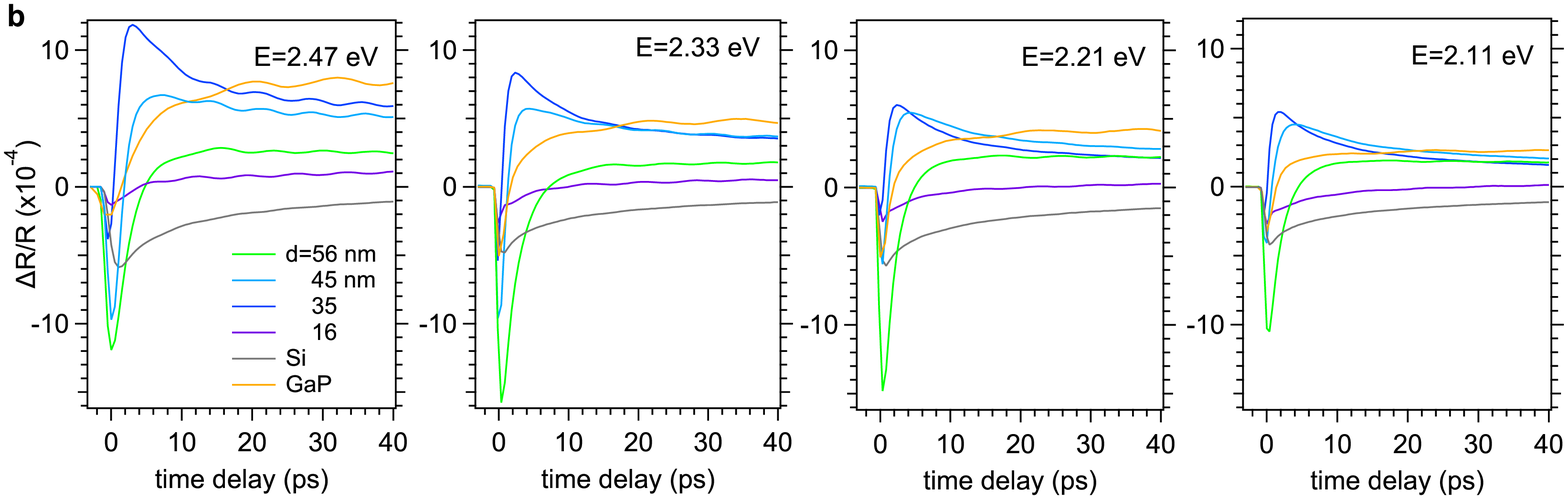}
\includegraphics[
width=0.9\textwidth]{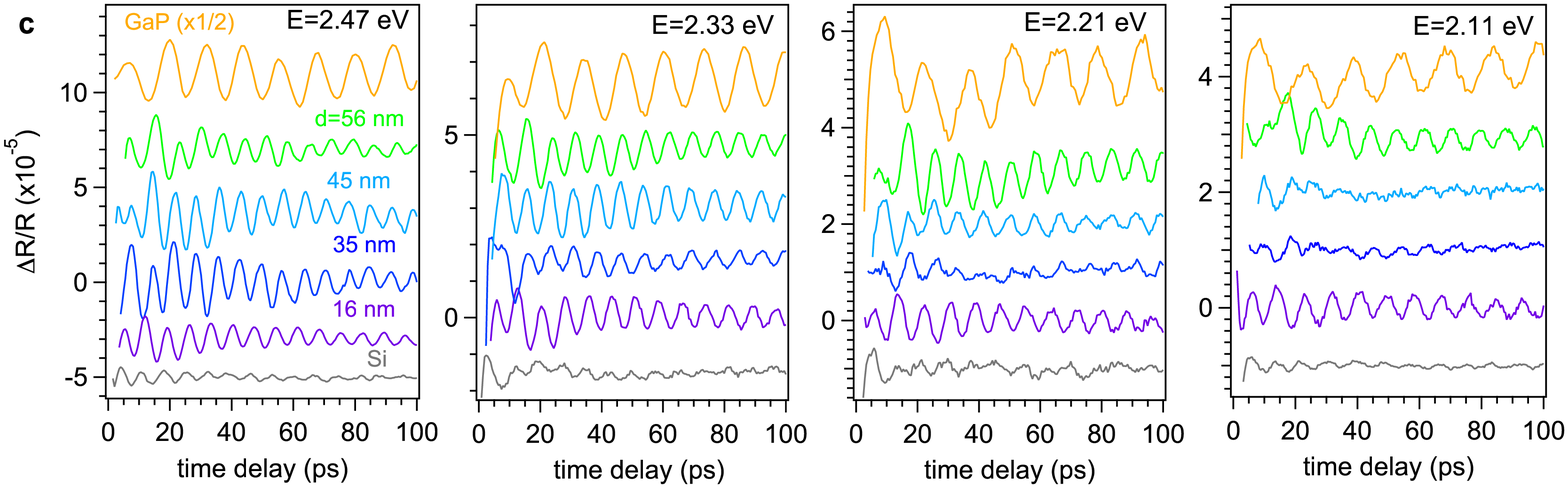}
\caption{\label{DRR_C} \textbf{a,b}: Transient reflectivity traces of GaP/Si(001) heterointerfaces with different GaP overlayer thicknesses $d$, together with those of bulk Si and GaP, pumped at 3.1 eV and probed at photon energies $E$.   Pump density is 90 $\mu$J/cm$^2$.  Panels \textbf{a} and \textbf{b} show the same traces in different horizontal and vertical scales.  \textbf{c}: The oscillatory part of the reflectivity traces in \textbf{a}.
}
\end{figure*}

The reflectivity traces for \emph{bulk} GaP and Si are also modulated periodically on much longer time scale, as shown in Fig.~\ref{TD400nm}b. 
These slow modulations are caused by the interference between the probe beams reflected from the propagating acoustic pulse and from the surface (Brillouin oscillation), as schematically shown in Fig.~\ref{Scheme}a, and their frequencies are given by $f_B =2nv/\lambda$=123 and 235 GHz for GaP and Si for normal incidence probe \cite{Thomsen1986, Ishioka2017PRB}.  Here $n$ is the refractive index  at the probe photon energy, as plotted in Fig.~\ref{alpha_n}b, and $v$ is the LA phonon velocity in the [001] direction  ($v^\textrm{GaP}$=5.847 and $v^\textrm{Si}$=8.4332 nm/ps \cite{Weil1968, McSkimin1964}).  Similar but less periodic modulations appear in the reflectivity traces for the GaP/Si interfaces, as shown Fig.~\ref{TD400nm}b. Their FT spectra, shown in Fig.~\ref{FT400nmL}a, feature both $f_B^\textrm{GaP}$ and $f_B^\textrm{Si}$ frequency-components, indicating an origin associated with acoustic pulses propagating in GaP and Si.  The frequent phase jumps in the quasi-periodic modulations can be qualitatively understood in terms of multiple acoustic pulses, generated at the top surface and at the interface as schematically shown in Fig.~\ref{Scheme}, passing the heterointerface and being reflected at the surface and interface.  A simple theoretical model neglecting the carrier transports within the semiconductors and across the heterointerface cannot reproduce the reflectivity modulations quantitatively, however \cite{Ishioka2017APL}. 

In addition to the quasi-periodic modulations, the reflectivity traces from the GaP/Si interfaces exhibit extra features seen as relatively sharp spikes, as indicated by arrows in Fig.~\ref{TD400nm}a \cite{Ishioka2017APL}.  These spikes are absent for bulk GaP and Si and are therefore characteristic of the heterointerface.  The spikes appear at delay times $t\sim d/v^\textrm{GaP}$, as plotted in Fig.~\ref{FT400nmL}b, implying that they are induced by the acoustic pulse generated at the GaP/Si interface and detected at the GaP/air surface, and/or vice versa.  The temporal width of the spikes, $\sim$0.5 ps, translates to a spatial extent of $\sim$3 nm or $\sim$10 atomic Ga-P layers, which is comparable with the intermixing of 7 atomic layers due to the pyramidal structures at the interface \cite{Beyer2016}.  The extremely narrow spatial extent of the acoustic pulse suggests a potential application  as an opto-acoustic transducer for high-resolution nano-seismology for Si-based structures.

\subsection{Two-color pump-probe measurements}

Figure~\ref{DRR_C}a shows the transient reflectivity changes of GaP/Si interfaces, together with those of bulk Si and GaP, pumped at 3.1 eV and probed at different photon energies $E$ in the visible range.  Like in the one-color measurements, $\Delta R/R$ comprises the non-oscillatory component associated with the carrier dynamics and the periodic modulation with sub-THz frequency that is indicative of the propagating acoustic pulse.  Unlike the one-color measurements, however, the durations of the pump and probe pulses in the two-color measurements are too long to excite and detect the coherent LO phonons of GaP and Si.  

The reflectivity traces of \emph{bulk} Si drop rapidly after photoexcitation and recover bi-exponentially with time constants of 7 and 65 ps that are independent of $E$.  For bulk GaP, the reflectivity traces show an initial rapid rise, followed by a slower increase and then a decay with the time constants of 11 and 290 ps being independent of $E$.  The shorter time constant is roughly in the range of the $L\rightarrow X$ scattering in Si \cite{Sangalli2015} and the $X_3\rightarrow X_1$ scattering in GaP \cite{Collier2013, Cavicchia1995}, whereas the slower decay is more likely to be due to the phonon-phonon scattering (lattice heating) \cite{Gunnella2016} and the surface recombination \cite{Sabbah2002}.  We note that the insensitivity of the time constants to $E$ suggests a minor contribution from the longitudinal diffusion of photoexcited carriers, since the probe depth $\alpha^{-1}$ depends critically on $E$ [Fig.~\ref{alpha_n}].

The reflectivity traces for the GaP/Si heterointerfaces at $d$=16 and 56 nm are somewhat similar to those of bulk Si and GaP, respectively,  in particular at relatively low $E$, as shown in Fig.~\ref{DRR_C}ab.  By contrast, those for $d$=35 and 45 nm rise almost instantaneously after photoexcitation and decay bi-exponentially, unlike bulk Si or GaP.  For all the heterointerfaces, the decay/rise on sub-50 ps time scale becomes faster with increasing $E$.  The observations indicate that the interfacial carrier dynamics dominate the reflectivity response.
	
\begin{figure}
\includegraphics[
clip,width=0.375\textwidth]{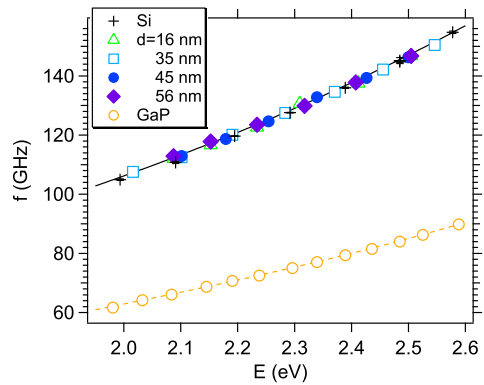}
\caption{\label{frequency} Frequency $f$ of the reflectivity modulation as a function of probe photon energy $E=hc/\lambda$.  Symbols represent experimental data.  The solid (dotted) curve represents $f=2nv/\lambda$ with $n$ and $v$ being the refractive index and the LA phonon velocity of Si (GaP).
}
\end{figure}

The sub-THz reflectivity oscillations of all the GaP/Si heterointerfaces, shown in Fig.~\ref{DRR_C}c,  in principle comprise only one frequency, $f_B^\textrm{Si}$ at given $E$, as shown in Fig.~\ref{frequency}.  Unlike in the one-color measurements, we see no clear sign of the modulation at $f_B^\textrm{GaP}$ except for the small wiggles for the first cycle or two.  This can be understood in terms of the poorer sensitivity of the visible probe light to the near surface region.  The periodic modulation at $f_B^\textrm{Si}$ dephases before the strain pulse goes out of the probe depth due to the broad bandwidth of the probe light \cite{Ishioka2017PRB}.
	
\begin{figure}
\includegraphics[
width=0.375\textwidth]{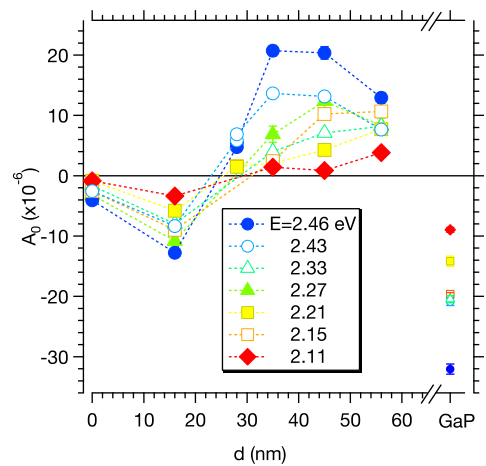}
\caption{\label{A_d} Initial amplitude $A_0$ of reflectivity modulation of GaP/Si interfaces as a function of GaP overlayer thickness $d$ measured at different probe photon energies $E$.
}
\end{figure}

Figure~\ref{A_d} plots the initial amplitude $A_0$, obtained by fitting the reflectivity oscillations for $t\gtrsim$10 ps to a damped harmonic function: 
\begin{equation}
g(t)=A_0 \cos(2\pi f_B^\textrm{Si}t+\phi), 
\end{equation}
as a function of the GaP thickness $d$.  We notice that  $A_0$ for bulk Si ($d=$0 nm) is negative, i.e., the reflectivity modulation is approximately proportional to $-\cos(2\pi f_B^\textrm{Si}t)$, at all $E$.  $A_0$ for the GaP/Si heterointerface for $d$=16 nm is also negative but its absolute value is significantly larger than that of bulk Si for all $E$.  With further increasing $d$, $A_0$ exhibits an apparently oscillatory behavior, with its sign flipped from negative for $d$=16 nm to positive for the larger $d$.  For bulk GaP, where the reflectivity is modulated at $f_B^\textrm{GaP}$, the initial amplitude $A_0$ is negative again.   The $d$-dependence cannot be explained simply by the absorption of the pump and probe lights by the GaP overlayer, because that would decrease $A_0$ monotonically with increasing $d$.  Nor can it be explained in terms of the $d$-dependent phase-shift of the pump light due to the different refractive index in GaP and Si, since the coherent acoustic phonons are not induced directly by the electric field of the pump light but through the photoexcited carriers in the present experimental scheme.  To understand the seemingly counterintuitive thickness-dependence, we consider a theoretical model of optical generation and detection of the acoustic pulse at the heterointerface in the following section.  

\section{discussion}

We start to theoretical model the optical detection of the coherent acoustic phonons by considering a probe light wave, which is described by $E_0(z,t)=E_e(t)e^{i(kz-\omega t)}$ in vacuum, incident on a semiconductor surface (without an overlayer) from the normal direction.  Once inside the semiconductor, whose complex refractive index is $\tilde{n_1}\equiv n_1+i\kappa_1$, the light wave is expressed by:
\begin{equation}\label{th1}
E_1(z,t)=t_{01}E_e(t)\exp[i(\tilde{k_1} z-\omega t)],
\end{equation}
with $\tilde{k_1}\equiv\tilde{n_1}k$ being the light wavevector in the semiconductor.  The transmission coefficients for the light incoming to and outgoing from the semiconductor are given by
\begin{equation}\label{th2}
t_{01}=\dfrac{2}{1+\tilde{n_1}};  \quad t_{10}=\dfrac{2\tilde{n_1}}{1+\tilde{n_1}} 
\end{equation}
The complex reflection coefficient (in the absence of the photoexcitation) is expressed by:
\begin{equation}\label{th3}
r_{01}=\dfrac{1-\tilde{n_1}}{1+\tilde{n_1}}
\end{equation} 
We assume that a separate pump light induces a change  $\delta\tilde{n_1}(k,z,t)$ in the refractive coefficient.   Since the reflected probe light crosses the semiconductor surface at $z=0$ twice and propagates in the semiconductor both ways, the corresponding change in the complex reflection coefficient is given by \cite{Cook2010, Liu2005}:
\begin{eqnarray}\label{th4}
\delta r&=&t_{01}t_{10}\left[-\dfrac{1}{2\tilde{n_1}}\int_{0}^{\infty}  e^{2i\tilde{k_1} z}\dfrac{\partial }{\partial z}\delta \tilde{n_1}(z,t)  dz\right]\nonumber\\
&=&\dfrac{-2}{(1+\tilde{n_1})^2}\int_{0}^{\infty}  e^{2i\tilde{k_1} z}\dfrac{\partial}{\partial z} \tilde{n_1}(z,t)  dz;\nonumber\\
\dfrac{\delta r}{r_{01}}&=&\dfrac{2}{\tilde{n_1}^2-1}\int_{0}^{\infty}  e^{2i\tilde{k_1} z}\dfrac{\partial}{\partial z}\delta \tilde{n_1}(z,t)  dz
\end{eqnarray}
The intensity of the reflected probe light in the absence of the pump light is given by $R_0=r_{01}r_{01}^*$.  The pump-induced change in the reflected probe intensity can be expressed by:
\begin{eqnarray}\label{th5}
\dfrac{\Delta R}{R_0}&\equiv&\dfrac{R-R_0}{R_0}\nonumber\\
&=&\dfrac{(r_{01}+\delta r)(r_{01}^*+\delta r^*)-r_{01}r_{01}^*}{r_{01}r_{01}^*}\nonumber\\
&\simeq& \dfrac{\delta r^*}{r_{01}^*}+\dfrac{\delta r}{r_{01}}=2\textrm{Re}\left(\dfrac{\delta r}{r_{01}}\right).
\end{eqnarray}
when $\delta r$ is small ($\delta r\ll r_{01}$).  By substituting eq.~(\ref{th4}) into (\ref{th5}) we obtain:
\begin{equation}\label{th6}
\dfrac{\Delta R}{R_0}\simeq\textrm{Re}\left[\dfrac{4}{\tilde{n_1}^2-1}\int_{0}^{\infty}  e^{2i\tilde{k_1} z}\dfrac{\partial}{\partial z}\delta \tilde{n_1}(z,t)  dz\right]
\end{equation}

Now we consider the change in the refractive index induced by a pump-induced strain $\eta(z,t)$:
\begin{equation}\label{th7}
\delta\tilde{n_1}(z,t)=\dfrac{\partial \tilde{n_1}}{\partial\eta}\eta(z,t)=\dfrac{1}{2\tilde{n_1}}\dfrac{\partial\epsilon}{\partial\eta}\eta(z,t)
\end{equation}
where $\epsilon\equiv\epsilon_r+i\epsilon_i=\tilde{n_1}^2$ is the complex dielectric constant.  We assume that the strain modulates the semiconductor bandgap and thereby the dependence of $\epsilon$ on the probe photon energy  $E$ \cite{Ishioka2017PRB}:
\begin{equation}
\epsilon(E,\eta)\simeq\epsilon(E-a_{cv}\eta).
\end{equation}
Here $a_{cv}\equiv-K(\partial E_g/\partial p)$ is the relative deformation potential coupling constant, $K$, the bulk modulous, $E_g$, the band gap, and $p$, the pressure.  Then we can approximately express the variation in the strain-induced dielectric constant at a fixed probe energy $E=hck$ by:
\begin{equation}\label{Th12}
\dfrac{\partial\epsilon}{\partial \eta}=\dfrac{\partial E}{\partial \eta}\dfrac{\partial\epsilon}{\partial E}\Big\vert_{E=hck}\simeq-a_{cv}\frac{\partial\epsilon}{\partial E}\Big\vert_{E=hck}
\end{equation}
and that in the refractive index by:
\begin{equation}
\delta\tilde{n_1}(z,t)\simeq-\dfrac{a_{cv}}{2\tilde{n_1}}\dfrac{\partial\epsilon}{\partial E}\Big\vert_{E=\hbar ck}\eta(z,t)
\end{equation}
By substituting this into eq. (\ref{th6}) we obtain
\begin{equation}\label{th8}
\dfrac{\Delta R}{R_0}\simeq\textrm{Re}\left[\dfrac{2a_{cv}}{\tilde{n_1}(\tilde{n_1}^2-1)}\dfrac{\partial \epsilon}{\partial E}\Big\vert_{E=\hbar ck}\int_{0}^{\infty}  e^{2i\tilde{k_1} z}\dfrac{\partial\eta}{\partial z}  dz\right]
\end{equation}
The generation of coherent acoustic phonons in bulk GaP and Si is dominated by the deformation potential electron-phonon coupling with photoexcited carriers \cite{Ishioka2017PRB}.  If we neglect the transport and recombination of photoexcited carriers and approximate the depth distribution of the carrier density by: 
\begin{equation}
N(t)=\begin{cases}
     0 & \text{for}\quad t<0\\
    \alpha_{pu} (1-R_{pu}) F e^{-\alpha_{pu} z} /E_{pu} & \text{for}\quad t\geq0
     \end{cases}
\end{equation}
the strain can be expressed as a linear function of the pump light fluence $F$ and $a_{cv}$ by \cite{Ishioka2017PRB}: 
\begin{eqnarray}
\eta(z,t)&=&\dfrac{\alpha_{pu} a_{cv}(1-R_{pu})F}{2 E_{pu}\rho v^2} \Biggl[ e^{-\alpha_{pu}(z+vt)}\left( e^{\alpha_{pu} v t}-1\right)^2\nonumber\\
&-&(vt-z)\left( e^{-\alpha_{pu}(z-vt)}+e^{\alpha_{pu}(z-vt)}\right)\Biggr]
\end{eqnarray}
Here $\alpha_{pu}, R_{pu}$, and $E_{pu}$ are the absorption coefficient, the reflectivity and the photon energy for the pump light.
This yields a photo-induced strain consisting of a step-like wave front that is generated at the surface ($z=0$) at $t=0$ and propagates into the depth direction with velocity $v$, and an elastic component that is only moderately dependent on $z$.  We therefore approximate the differential strain with a delta function:
\begin{equation}\label{th9}
\dfrac{\partial\eta}{\partial z}\simeq B\delta(z=vt)
\end{equation}
with a real amplitude $B$, and obtain:
\begin{eqnarray}
\dfrac{\Delta R}{R_0}&\simeq&\textrm{Re}\left[\dfrac{2a_{cv}B}{\tilde{n_1}(\tilde{n_1}^2-1)}\dfrac{\partial \epsilon}{\partial E}\Big\vert_{E=\hbar ck}\int_{0}^{\infty}e^{2i\tilde{k_1} z}\delta(z=vt)  dz\right]\nonumber\\
&=&\textrm{Re}\left[\dfrac{2a_{cv}B}{\tilde{n_1}(\tilde{n_1}^2-1)}\dfrac{\partial \epsilon}{\partial E}\Big\vert_{E=\hbar ck}e^{2i\tilde{k_1} vt}  \right].
\end{eqnarray}
 Since $n_1\gg\kappa_1$ for the visible light in Si and GaP \cite{Aspnes1983},  we can safely neglect the imaginary part of $\tilde{n_1}$ and approximate the variation in the reflected probe intensity by:
\begin{eqnarray}\label{th10}
\dfrac{\Delta R}{R_0}&\simeq&\dfrac{2a_{cv}B}{n_1(n_1^2-1)}\textrm{Re}\left[\dfrac{\partial \epsilon}{\partial E}\Big\vert_{E=\hbar ck}e^{2ik_k vt}  \right]\nonumber\\
&=&\dfrac{2a_{cv}B}{n_1(n_1^2-1)}\Big\vert\dfrac{\partial\epsilon}{\partial E}\Big\vert\textrm{Re}\left[e^{i(2k_1 vt+\phi)}\right],
\end{eqnarray}
with $k_1\equiv n_1k$ and 
\begin{eqnarray}
\Big\vert\dfrac{\partial\epsilon}{\partial E}\Big\vert^2&\equiv&\dfrac{\partial \epsilon_r}{\partial E}\Big\vert_{E=\hbar ck}^2+\dfrac{\partial\epsilon_1}{\partial E}\Big\vert_{E=\hbar ck}^2;\nonumber\\
\tan\phi&\equiv&\dfrac{\partial \epsilon_r/\partial E}{\partial \epsilon_i/\partial E}\nonumber
\end{eqnarray}
Eq.~(\ref{th10}) describes the periodic modulation in the reflected probe intensity  (Brillouin oscillation) at a frequency $f_B=2k_1v/2\pi=2n_1v/\lambda$ for bulk semiconductors. 

\begin{figure}
\includegraphics[
width=0.35\textwidth]{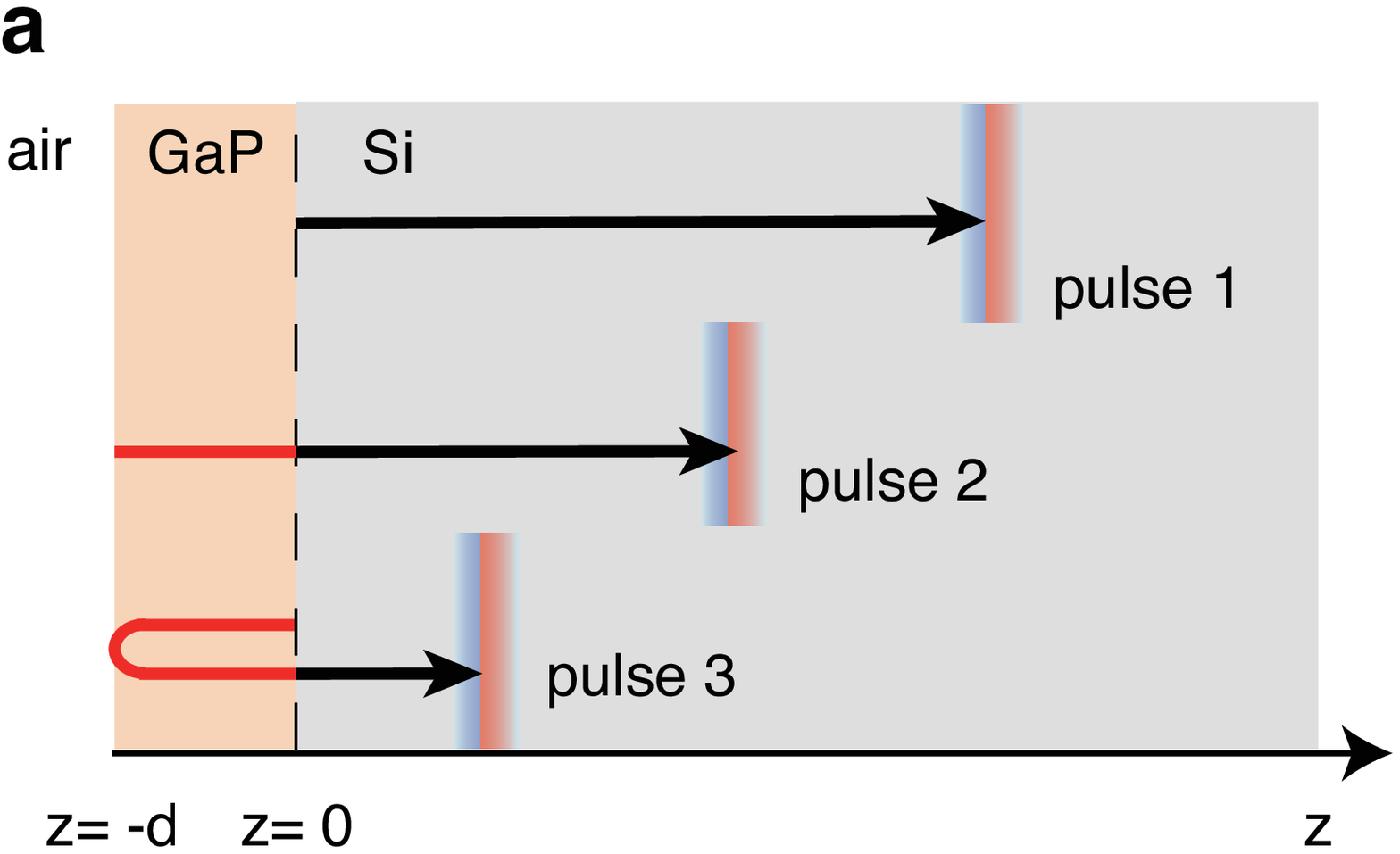}
\includegraphics[
width=0.475\textwidth]{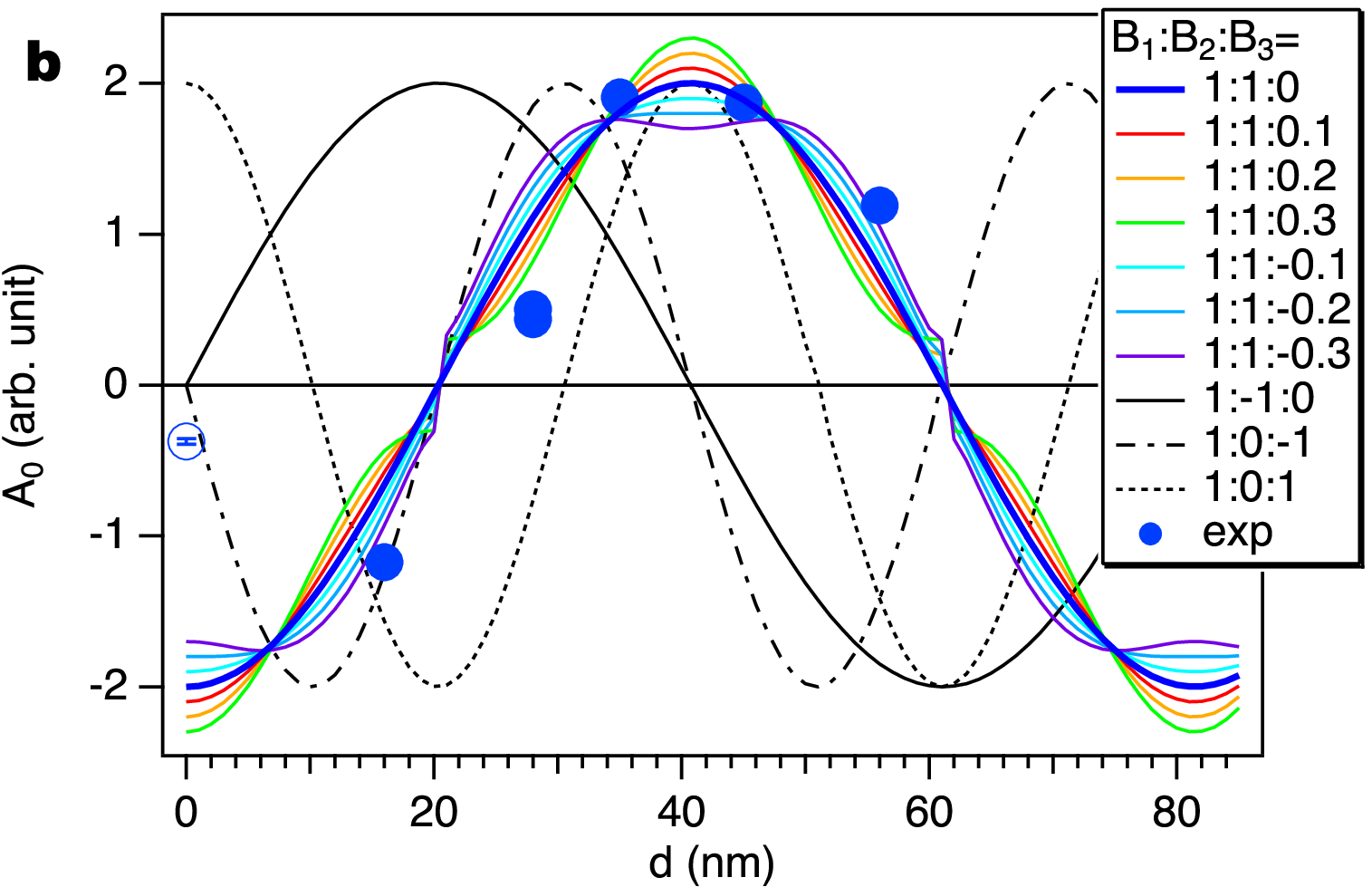}
\caption{\label{calc_A} \textbf{a}: Schematic illustration showing the positions of acoustic pulse at time delays $t\geq2d/v^\textrm{GaP}$.  \textbf{b}: Calculated $d$-dependence of $A_0$ at probe photon energy $E$=2.46 eV with different ratios between amplitudes $B_1, B2$ and $B_3$ in eq.~(\ref{th22}).  Filled and open circles represent the experimental $A_0$ of Si with and without GaP the overlayer measured at $E$=2.46 eV.
}
\end{figure}

The strain-induced reflectivity change for the GaP/Si interface has a much more complex expression, because the acoustic pulses can be generated also at the heterointerface, and because they can be reflected when they reach the boundaries of the GaP layer.  At long time delays of $t\geq2d/v^\textrm{GaP}$, however, the strain pulses are all located in the Si substrate and propagating in the depth direction, as schematically shown in Fig.~\ref{calc_A}a, if we neglect the small ($\sim$10 \%) reflection at the heterointerface.  In this case the differential strain can be simplified into an expression:
\begin{equation}\label{th11}
\dfrac{\partial\eta}{\partial z}\simeq B_1\delta(z=z_1)+B_2\delta\left(z=z_2\right)+B_3\delta\left(z=z_3\right)\end{equation}
with
\begin{eqnarray}\label{th12}
z_1(t)&=&v^\textrm{Si}t;\nonumber\\
z_2(t)&=&v^\textrm{Si}(t-d/v^\textrm{GaP}); \nonumber\\
z_3(t)&=&v^\textrm{Si}(t-2d/v^\textrm{GaP})
\end{eqnarray}
being the positions of pulse 1, 2 and 3 with respect to the GaP/Si interface.  We then obtain the approximate expression for the reflectivity modulation from the heterointerface:
\begin{eqnarray}\label{th13}
\dfrac{\Delta R}{R_0}&\simeq&\dfrac{2a_{cv}}{n_1(n_1^2-1)}\Big\vert\dfrac{\partial\epsilon}{\partial E}\Big\vert\textrm{Re}\Biggl[B_1\exp i{(2k_1v^{Si}t+\phi)}\nonumber\\
&+&B_2\exp i\left[2k_1 v^{Si}\left(t-\frac{d}{v^{GaP}}\right)+\phi\right]\nonumber\\
&+&B_3\exp i\left[2k_1 v^{Si}\left(t-\frac{2d}{v^{GaP}}\right)+\phi\right]\Biggr]\end{eqnarray}
By using a polar form:
\begin{eqnarray}
B_1&+&B_2\exp \left(\frac{-2ik_1v^{Si}d}{v^{GaP}}\right)+B_3\exp \left( \frac{-4ik_1v^{Si}d}{v^{GaP}}\right)\nonumber\\
&\equiv& \vert A_0\vert e^{i\psi}
\end{eqnarray}
the change in the reflected probe intensity can be expressed by:
\begin{eqnarray}
\dfrac{\Delta R}{R_0}&\simeq&\dfrac{2a_{cv}\vert A_0\vert}{n_1(n_1^2-1)}\Big\vert\dfrac{\partial\epsilon}{\partial E}\Big\vert\textrm{Re}\left[\exp {(2ik_1v^{Si}t+\phi+\psi)}\right]\nonumber\\
&=&\dfrac{2a_{cv}\vert A_0\vert}{n_1(n_1^2-1)}\Big\vert\dfrac{\partial\epsilon}{\partial E}\Big\vert\cos (2k_1v^{Si}t+\phi+\psi).
\end{eqnarray}
Here the amplitude and the phase are defined by:
\begin{eqnarray}\label{th22}
\vert A_0\vert^2&\equiv&\left[B_1+B_2\cos \left(\frac{2k_1 v^{Si}d}{v^{GaP}}\right)+B_3\cos \left(\frac{4k_1 v^{Si}d}{v^{GaP}}\right)\right]^2\nonumber\\
&+&\left[B_2\sin \left(\frac{2k_1 v^{Si}d}{v^{GaP}}\right)+B_3\sin \left( \frac{4k_1 v^{Si}d}{v^{GaP}}\right)\right]^2;
\end{eqnarray}
\begin{equation}
\tan\psi\equiv\dfrac{B_1+B_2\cos \left(\frac{2k_1v^{Si}d}{v^{GaP}}\right)+B_3\cos \left(\frac{4k_1 v^{Si}d}{v^{GaP}}\right)}{B_2\sin \left(\frac{2k_1v^{Si}d}{v^{GaP}}\right)+B_3\sin \left(\frac{4k_1 v^{Si}d}{v^{GaP}}\right)}\nonumber
\end{equation}

Figure~\ref{calc_A}b compares the calculated $d$-dependence of $A_0$ at $E$=2.46 eV with the experiment.  We see that eq.~(\ref{th22}) reproduces the experimental oscillatory behavior reasonably well when we consider the interference arising only from the pulses 1 and 2 by putting $B_1:B_2:B_3=1:1:0$, but not with $1:-1:0$.  In both cases, adding small positive or negative $B_3$ does not change the calculated $d$-dependence drastically.  If we consider the interference arising only from the pulses 1 and 3 by putting $B_1:B_2:B_3=1:0:\pm1$, the oscillation with $d$ has twice the frequency and does not match the experiment.  The results confirm that the interference between the reflections from multiple acoustic pulses is responsible for the oscillatory $d$-dependence, with the contribution from pulse 2 as important as from pulse 1.  The calculations suggest that the pulses 1 and 2 have the amplitudes with the same sign, which gives us a hint in determining which the satellite valley is most relevant in the acoustic phonon generation, since $a_{cv}$ for different valleys can have different signs \cite{Cardona1987, vdWalle1989}.  The contribution from pulse 3 is small in comparison, $\vert B_3/B_1\vert<0.3$, which can be attributed to the partial loss when it is reflected by the GaP surface and transmits across the heterointerface.  
 
\section{summary}

We have investigated the ultrafast electron-phonon coupling at lattice-matched heterointerfaces of GaP/Si(001) in the form of coherent optical and acoustic phonons.  The screening of the polar LO phonons of GaP with photoexcited plasmas depends critically on the GaP layer thickness, which is attributed to the quasi-2D character of the plasmons confined in the thin GaP layers.  The presence of the GaP overlayer on top enhances the acoustic pulse that propagates in Si through the larger deformation potential electron-phonon coupling in GaP.  Interference between the probe lights reflected by multiple acoustic pulses generated at the GaP/Si interface \emph{and} at the GaP top surface leads to the reflectivity modulation whose amplitude depends on the GaP overlayer thickness.  We have thereby demonstrated the ultrafast electron-phonon couplings that are characteristic of a semiconductor heterostructure with an abrupt, almost defect-free interface.  The insight obtained in the present study will stimulate the development of theoretical simulations that better describe the optical generation and detection of acoustic phonons at the GaP/Si and other semiconductor heterointerfaces.

\begin{acknowledgments}
This work is partly supported by the Deutsche Forschungsgemeinschaft through SFB 1083 and HO2295/8, as well as  the Air Force Office of Scientific Research under Award No. FA9550-17-1-0341 (Stanton) and NSF grant DMR-1311845 (Petek).  
\end{acknowledgments}

\bibliographystyle{iopart-num}
\bibliography{GaPSi2018}

\end{document}